\RequirePackage{pdf14}
\documentclass[3p,times,procedia]{elsarticle}

\usepackage{nupha_ecrc}
\usepackage{graphicx}

\usepackage[    bookmarks,
                 bookmarksopen = true,
                 bookmarksnumbered = true,
                 breaklinks = true,
                 linktocpage,
                 pagebackref,
                 colorlinks = true,
                 linkcolor = blue,
                 urlcolor  = blue,
                 citecolor = purple,
                 anchorcolor = green,
                 hyperindex = true,
                 hyperfigures]
                 {hyperref}

\volume{00}

\firstpage{1}

\journalname{Nuclear Physics A}

\runauth{H.~Li and L.~Yan}

\jid{nupha}

\jnltitlelogo{Nuclear Physics A}

\usepackage{amssymb}

\usepackage[figuresright]{rotating}

\def\x{{\bm x}}

\newcommand\nda{\end{align}}
\def\Eq#1{Eq.~(\ref{#1})}
\def\Fig#1{Fig.~\ref{#1}}

\def\bra{\left\langle}
\def\ket{\right\rangle}
\def\E{\mathcal{E}}

\def\t{\tilde}

\advance\parskip 1.9pt
\advance\voffset -0.2in

\def\be{\begin{equation}}
\def\ee{\end{equation}}
\def\bg{\begin{eqnarray}}
\def\nd{\end{eqnarray}}

\begin{document}

\begin{frontmatter}



\dochead{XXVIIIth International Conference on Ultrarelativistic Nucleus-Nucleus Collisions\\ (Quark Matter 2019)}

\title{Longitudinal fluid response and pseudorapidity dependent flow in relativistic heavy-ion collisions}


\author{Hui Li}
\author{Li Yan}
\ead{cliyan@fudan.edu.cn}

\address{Key Laboratory of Nuclear Physics and Ion-Beam Application (MOE) \& Institute of Modern Physics\\
Fudan University, 220 Handan Road, 200433, Yangpu District, Shanghai, China}

\begin{abstract}
We study the pseudorapidity dependent hydrodynamic response in heavy-ion collisions.
A differential hydrodynamic relation is obtained for elliptic flow.  
Using event-by-event simulations of 
3+1D MUSIC, with initial conditions generated via a multi-phase transport (AMPT) model, 
the differential response relation is verified. Based on the response relation, 
we find that the two-point correlation of elliptic flow in 
pseudorapidity are separated into the fluid response and the two-point correlation of initial eccentricity. 
\end{abstract}


\end{frontmatter}


\section{
Introduction
 }
 The fluidity of the quark-gluon plasma (QGP) created in heavy-ion collisions
has been discovered through the measurements of flow harmonics in multi-particle correlations. 
These flow harmonics can be understood as 
the fluid response to the decomposed azimuthal modes associated with
the initial state geometrical deformations. For instance, it is noticed 
that the eccentricity of the initial density profile, $\E_2$, is linear to the elliptic flow $V_2$~\cite{Ollitrault:1992bk}.
While this linear relation has been well studied 
by relativistic hydrodynamics~\cite{Qiu:2011iv,Niemi:2012aj,Noronha-Hostler:2015dbi}, 
a pseudorapidity dependent hydrodynamic response relation between $V_2$ and $\E_2$ is absent in the community, 
until some recent studies~\cite{Li:2019eni,Franco:2019ihq}.
In the current proceeding, we generalize 
the linear response relation for the second
flow harmonics, to a pseudorapidity dependent hydrodynamic response. With event-by-event simulations of 3+1D MUSIC with respect to initial condition from AMPT, 
the pseudorapidity dependent response relation is confirmed. 
Given the pseudorapidity dependent hydrodynamic
response, we are able to explore the relation between the two-point correlation of elliptic flow in 
pseudorapidity and that of initial eccentricity. 

\section{
Framework
 }
To study the pseudorapidity dependent hydrodynamic response, we generalize the linear 
 response relation  as
\be
\label{eq:resp}
V_2(\zeta) = \int_{-\infty}^\infty d\xi \; G(\zeta-\xi) \E_2(\xi)\,,
\ee
where $\zeta$ is the pseudorapidity and $\xi$ is the space-time rapidity. Although the response function $G(\zeta-\xi)$ implies a boost invariant background, 
the broken boost invariant symmetry in realistic heavy-ion collisions can be accounted for by perturbations. In \Eq{eq:resp}, the pseudorapidity dependent flow is defined according to the Fourier decomposition of 
the particle emission probability in heavy-ion collisions, $P(\phi_p, \zeta)$
\be
P(\phi_p, \zeta) = \frac{1}{2\pi} \sum_{n=-\infty}^\infty V_n(\zeta) e^{-in \phi_p}\,,\qquad
{\mbox{integrated flow: }}V_n = \int_{-\infty}^\infty d\zeta V_n(\zeta)\,.
\ee
 With respect to initial energy density $\rho(\vec x_\perp,\xi)$, the initial eccentricity is defined as
$ \E_2(\xi) = -\int d^2 \vec x_\perp \; \rho(\vec x_\perp,\xi) (x+iy)^2/
 \int d\xi d^2 \vec x_\perp \; \rho(\vec x_\perp,\xi) |x+iy|^2\,$. Up to integration over $\xi$, one finds that $\E_2=
 \int_{-\infty}^\infty \E_2(\xi)$ is the standard eccentricity considered in literature.  It is also worth mentioning that, asymptotically, $\E_2(\xi)\to0$ when $|\xi|\to\infty$, which would allow one to ignore boundary corrections from
 infinities when deriving \Eq{eq:linear_eta}.
 
To identify the response function, it is
advantageous to work 
through a Fourier transformation, namely, for the elliptic flow
$\t V_2(k) = \int_{-\infty}^\infty d\zeta \;V_2(\zeta) e^{-ik\zeta}$ and 
initial eccentricity $\t \E_2(k) =\int_{-\infty}^\infty d\xi \;\E_2(\xi) e^{-ik\xi}$.
In terms of the wave-number $k$, \Eq{eq:resp} becomes,
\be
\label{eq:lineark}
\t V_2(k)=\t G(k) \t \E_2(k)\,.
\ee
In the small wave number 
limit with $|k|\ll k^*$, corresponding to the hydrodynamic regime, the response function can be expanded in series of $k$. Up to the second order, the expansion is
\be
\label{eq:expk}
\t G(k)= G_0 + ikG_1 - k^2 G_2 + O(k^3)\,,
\ee
which in the $\zeta$-space amounts to 
\be
\label{eq:vn_eta}
V_2(\zeta) = G_0 \E_2(\zeta) + G_1 \frac{d\E_2(\zeta)}{d\zeta} + G_2 \frac{d^2 \E_2(\zeta)}{d \zeta^2}
+ O\left(\frac{d^3}{d\zeta^3}\right)\,.
\ee 
Note that odd order $G_n$ vanish owing to the parity symmetry in the background.
To obtain these expansion coefficients $G_n$'s, 
we define new sets of flow variables and
initial eccentricity variables weighted with powers of $\zeta$ (or $\xi$):
$V^{(n)}_2=\int d\zeta \zeta^n V_2(\zeta)/n!\,$ and 
$\E^{(n)}_2= \int d\xi \xi^n \E_2(\xi)/n!\,.$
Using integration by parts repeatedly, with these new variables 
the generalized linear response relation \Eq{eq:resp} can be rewritten as 
\be
\label{eq:linear_eta}
V_2^{(n)}=\sum_{i=0}^n (-1)^i G_i \E_2^{(n-i)}\,.
\ee
Note that the leading order relation $V_2^{(0)}=G_0 \E_2^{(0)}$ is the familiar linear response 
relation of elliptic flow, with the response coefficients being calculated in event-by-event hydrodynamic simulations as~\cite{Noronha-Hostler:2015dbi}:
$
G_0 = \bra V_2^{(0)} \E_2^{(0)*}\ket\Big/\bra \E_2^{(0)}\E_2^{(0)*}\ket
$, 
where and in the following the angular brackets indicate event average.
Following a similar procedure,  a set of linear relations can be realized
between $\E_2^{(0)}$ and $V_2^{(n)}-\sum_{i=0}^{n-1} (-1)^{i}G_i \E_2^{(n-i)}$, and
$G_n$ can be calculated recursively. More details on these higher coefficients can be found in Ref.\cite{Li:2019eni}.

The pseudorapidity dependent response relation \Eq{eq:resp} 
is non-local, which implies that the generation of
$V_2$ at one pseudorapidity receives contributions from other 
space-time rapidities.   
This effect can be shown in the analysis of two-point correlations. 
We define
\be
\label{eq:width_v2}
\bra (\Delta \zeta)^2\ket
\equiv
\frac{\int d\zeta d \zeta' \bra V_2(\zeta) V_2^*(\zeta')\ket(\zeta'-\zeta)^2 }
{\int d\zeta d\zeta' \bra V_2(\zeta) V_2^*(\zeta')\ket},
\ee
to characterize the length of the two-point correlation measured via elliptic flow
at different pseudorapidities. 
With the response relation derived in \Eq{eq:linear_eta}, 
it can be proved that 
\be
\label{eq:width_g2g0}
\bra (\Delta \zeta)^2\ket=\bra (\Delta \xi)^2\ket + 4G_2/G_0\,.
\ee
The length of the initial state eccentricity two-point correlation  
$\bra (\Delta \xi)^2\ket$ is defined according to \Eq{eq:width_v2} through 
$\E_2(\xi)$.

\section{Results and discussion}
\begin{figure}
  \begin{center}
  \includegraphics[width=0.4\textwidth, trim={0.1cm 0.05cm 0.03cm 0.03cm},clip]{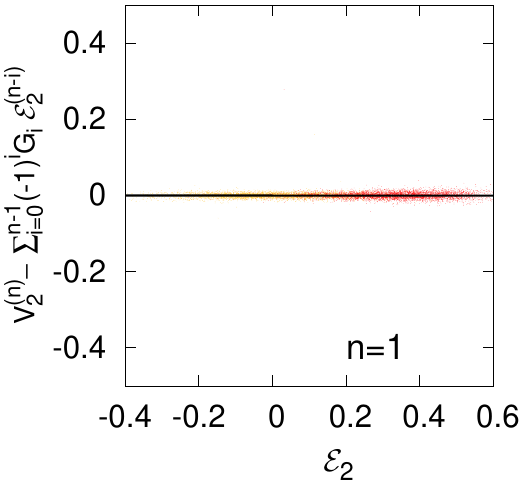}
  \includegraphics[width=0.4\textwidth, trim={0.1cm 0.05cm 0.05cm 0.03cm},clip]{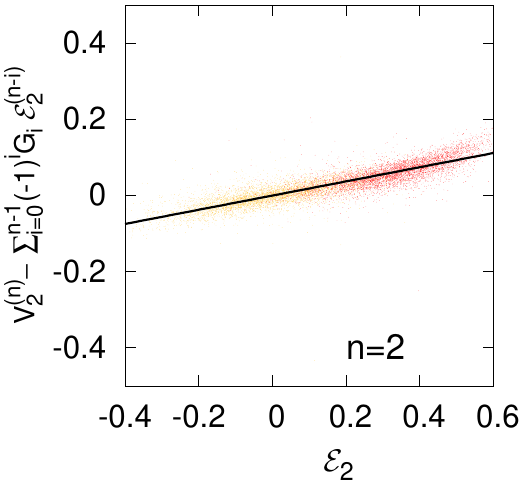}
 
  \caption{Linear relations respectively from the real part (red points) and the imaginary part (yellow points) for $n=1$ and $n=2$, from hydro simulations with constant $\eta/s=0.08$. Slopes of solid lines are determined from the values of $G_1$ and $G_2$ respectively. }
  \label{fig:Gn}
  \end{center}
 \end{figure}

To verify the pseudorapidity dependent response relation, we perform event-by-event 
hydrodynamic simulations,
for the Pb-Pb collision system with $\sqrt{s_{NN}}=2.76$ TeV at the LHC, within centrality class 30-40\%.
The 3+1D MUSIC ~\cite{Schenke:2010nt,Schenke:2010rr} is used with respect to random 3D initial conditions generated by the 
AMPT model~\cite{Lin:2004en}. The initial density profile is obtained similarly as 
in~\cite{Li:2017slc}. 
The pseudorapidity dependent elliptic flow $V_2(\zeta)$ is calculated from thermal pions. 
Given the results from hydrodynamic simulations, the linear relation 
between $\E_2^{(0)}$ and $V_2^{(n)}-\sum_{i=0}^{n-1} (-1)^{i}G_i \E_2^{(n-i)}$ can be verified,
which further determines the constant response coefficients as the slope.
\Fig{fig:Gn} shows the results from
the hydrodynamic simulations of approximately 5000 events, with a constant $\eta/s=0.08$. Only the case of
$n=1$ and $n=2$ are presented, but they are representative to illustrate the linear relations, and the fact that odd order $G_n$'s vanish.
The absolute values of even order $G_n$'s increase exponentially, which 
implies a finite radius of convergence, $k^*$, of the hydrodynamic gradient expansion in \Eq{eq:expk}. More detailed 
discussions on the convergence behavior can be found in Ref.\cite{Li:2019eni}.

\begin{figure}
  \begin{center}
  \includegraphics[width=0.4\textwidth, trim={0.1cm 0.05cm 0.05cm 0.03cm},clip]{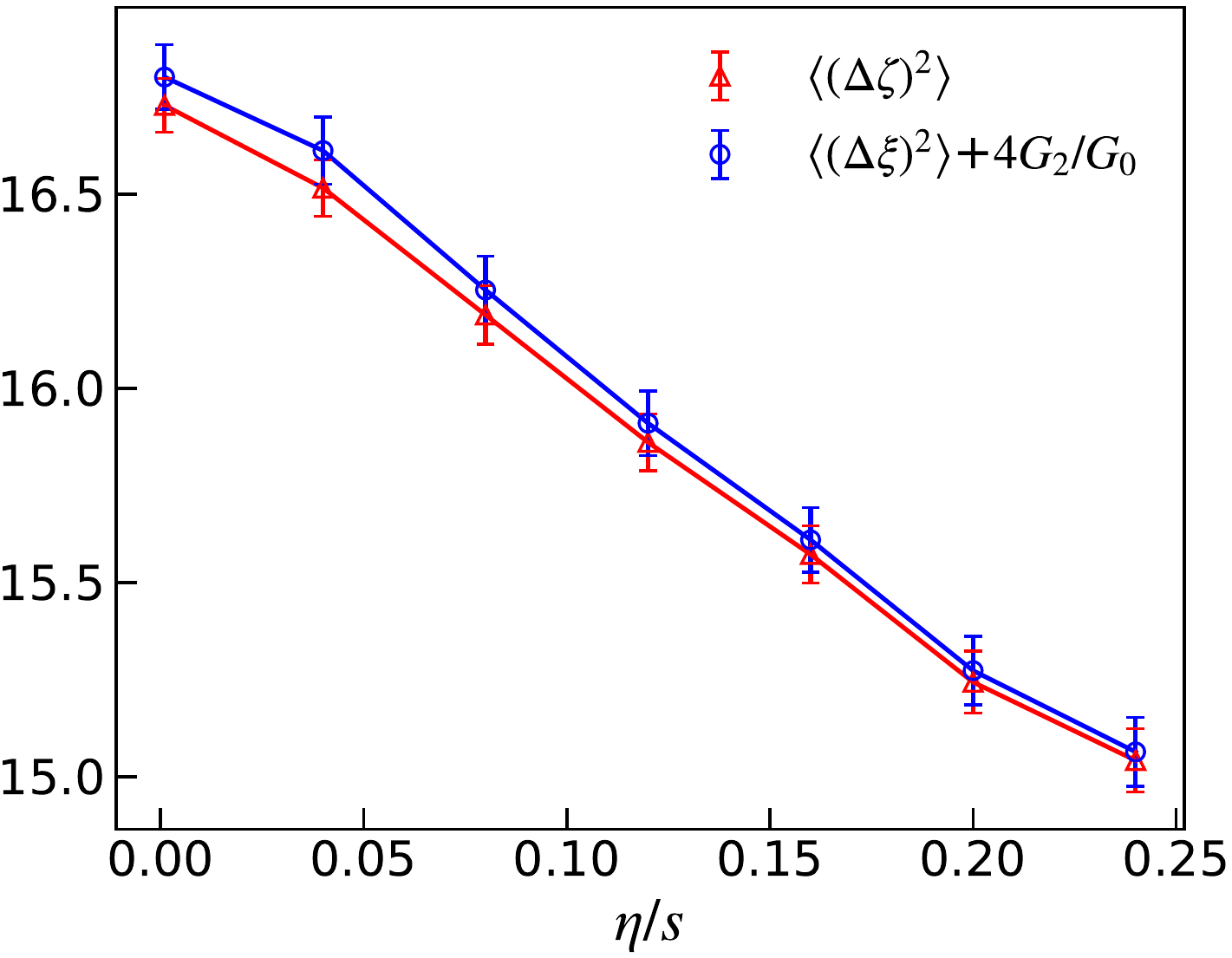}
 
  \caption{Two-point correlation length of $V_2$ and the 
expectations from \Eq{eq:width_g2g0}, as a function of $\eta/s$,
where the initial two-point correlation length is $\bra (\Delta \xi)^2\ket=12.81 \pm 0.08$.
Figure taken from Ref. \cite{Li:2019eni}.}
  \label{fig:2p}
  \end{center}
 \end{figure}

\Fig{fig:2p} shows the numerical results of about 1000 events from hydrodynamic simulations with different 
values of $\eta/s$ from 0.001 to 0.2. The obtained length of the two-point correlation of $V_2$ defined in \Eq{eq:width_v2} is plotted as a function
of $\eta/s$, which agrees with the expectation from  
\Eq{eq:width_g2g0} within statistical errors. This indicates that the \emph{increase} of the two-point correlation length in elliptic flow
comparing to that in the initial eccentricity is purely an effect of fluid dynamics.
One can also see that the two-point correlation at the final stage is reduced as the increase of $\eta/s$.
This can be understood as a direct consequence of sound propagation, reflected in the ratio
$G_2/G_0$~\cite{Kapusta:2011gt}.
Since sound propagation is damped as a result of fluid dissipation, the two-point correlation at the final stage is reduced.

In summary, we have derived the differential formulation of pseudorapidity dependent hydrodynamic response which is different from some previous studies (cf.~\cite{Bozek:2015bha,Sakai:2018sxp}). 
The formulation is expected to be applicable to other flow harmonics as well, when nonlinear response is not dominant in flow generation~\cite{Noronha-Hostler:2015dbi,Teaney:2012ke}.
Through the event-by-event simulations of 
3+1D MUSIC, with initial conditions generated via the AMPT model,  
the differential response relation is verified. We also find that the \emph{increase} of the two-point correlation of elliptic flow 
comparing to that in the initial eccentricity is purely an effect of fluid dynamics.

\section{Acknowledgements}
This work is supported by National Natural Science Foundation of China (NSFC) 
under Grant No. 11975079 (LY) and China Postdoctoral Science
Foundation under Grant No. 2019M661333 (HL).




\bibliographystyle{elsarticle-num}
    \setlength{\bibsep}{1pt}
    \linespread{1}\selectfont
    \bibliography{references.bib}







\end{document}